\documentclass[aps,pra,letterpaper,10pt,twocolumn,showpacs,superscriptaddress,notitlepage]{revtex4-2}

\usepackage[utf8]{inputenc}
\usepackage{amsfonts,amssymb,amsmath,amsthm}
\usepackage{url}
\usepackage{dsfont}
\usepackage{bbm}
\usepackage[usenames,dvipsnames]{xcolor}
\usepackage{xcolor}
\usepackage{nicefrac}
\usepackage{setspace}
\usepackage{tabularx,booktabs}
\usepackage{makecell}
\usepackage{comment}
\usepackage[normalem]{ulem}
\usepackage{siunitx}
\usepackage{physics}
\usepackage{comment}
\usepackage{graphicx}
\usepackage[colorlinks=true, citecolor=blue,urlcolor=blue,linkcolor=blue,filecolor=black]{hyperref}
\renewcommand{\selectlanguage}[1]{}

\def\virgolette #1{``#1"}

\usepackage{todonotes}

\begin{document}

\title{Hybrid encoder for discrete and continuous variable QKD}

\author{Mattia Sabatini}
\affiliation{Dipartimento di Ingegneria dell'Informazione, Universit\`a degli Studi di Padova, via Gradenigo 6B, 35131 Padova, Italy}

\author{Tommaso Bertapelle}
\affiliation{Dipartimento di Ingegneria dell'Informazione, Universit\`a degli Studi di Padova, via Gradenigo 6B, 35131 Padova, Italy}

\author{Paolo Villoresi}
\affiliation{Dipartimento di Ingegneria dell'Informazione, Universit\`a degli Studi di Padova, via Gradenigo 6B, 35131 Padova, Italy}
\affiliation{Padua Quantum Technologies Research Center, Università degli Studi di Padova, via Gradenigo 6A, 35131 Padova, Italy}

\author{Giuseppe Vallone}
\affiliation{Dipartimento di Ingegneria dell'Informazione, Universit\`a degli Studi di Padova, via Gradenigo 6B, 35131 Padova, Italy}
\affiliation{Padua Quantum Technologies Research Center, Università degli Studi di Padova, via Gradenigo 6A, 35131 Padova, Italy}

\author{Marco Avesani}
\thanks{marco.avesani@unipd.it}
\affiliation{Dipartimento di Ingegneria dell'Informazione, Universit\`a degli Studi di Padova, via Gradenigo 6B, 35131 Padova, Italy}
\affiliation{Padua Quantum Technologies Research Center, Università degli Studi di Padova, via Gradenigo 6A, 35131 Padova, Italy}


\begin{abstract}
Quantum key distribution (QKD) is emerging as a cutting-edge application of quantum technology, gradually integrating into the industrial landscape.
Many protocols employing discrete or continuous variables have been developed over time.
Whereas the firsts usually excel in covering longer distances, the seconds are typically superior in producing higher secret key rates at short distances.
Present efforts aim to create systems that can exploit both these strengths, foreseeing the future challenge regarding the realization of a quantum network consisting of multiple and heterogeneous interconnected nodes.
Within such a context, a possible solution are devices able to efficiently toggle between discrete and continuous variable working modes with hybrid quantum state encoders.
Therefore, this study presents a new hybrid encoder based on an iPOGNAC modulator, ensuring compatibility with Discrete Variable (DV) and Continuous Variable (CV) QKD systems that can be assembled entirely with commercial-off-the-shelf components.
The proposed scheme is the first supporting DV polarization protocols, thus making it an appealing candidate for space nodes of a future quantum network, given that polarization-based protocols are well suited for space links.
\end{abstract}

\maketitle


\section{Introduction}
One of the main goals of cryptography is to enable confidential and secure communication between two parties over an un-trusted channel.
To accomplish this, most existing protocols depend on computational security, meaning that they rely on mathematical problems believed to be unfeasible to solve by computers.
However, this belief remains unproven, and it is conceivable that an efficient algorithm, whether classical or quantum, may exist to tackle the latter.
Quantum Key Distribution (QKD) \cite{scarani_security_2009, gisin_quantum_2002, xu_secure_2020} offers a way to overcome the latter limitation by harnessing the principles of quantum mechanics.
The approach enables the development of unconditionally secure protocols that can withstand adversaries possessing limitless classical or quantum computational power.

According to the degrees of freedom of the quantum phenomenon used, QKD can be categorized as Discrete Variables (DV) or Continuous Variables (CV) \cite{weedbrook_gaussian_2012, laudenbach_continuous-variable_2018, pirandola_advances_2020, zhang_continuous-variable_2024}.
The former usually exploits single photons, such as polarization and time-bin, whereas CV uses the quadratures of the quantized electromagnetic field.
The BB84 protocol pioneered DV-QKD \cite{bennett_quantum_2014}.
It relies on single photon sources and non-orthogonal states for information encoding.
Over time, it was refined and underwent several improvements, such as decoy states to counter photon number-splitting attacks \cite{brassard_limitations_2000} and adopting fewer states to ease practical implementation \cite{grunenfelder_simple_2018}.
For CV-QKD, GG02 is the protocol of reference \cite{grosshans_continuous_2002}.
It employs Gaussian states to encode the secret key and coherent detection to measure them.
However, significant practical limitations hinder its adoption, such as requiring a continuous Gaussian modulator, low reconciliation efficiency, and computationally demanding error correction procedures \cite{almeida_secret_2021}.
These challenges can be mitigated using coherent state discrete constellation modulation protocols \cite{leverrier_unconditional_2009,denys_explicit_2021}.
However, this comes at the price of limiting the number of secret bits that can be encoded per symbol, affecting thus the Secret Key generation Rate (SKR) when compared to the original GG02.

Discrete and continuous protocols offer different advantages over the other.
DV-QKD is known for its more mature security proofs and ability to cover longer distances, reaching a record of $421~\si{\kilo\meter}$ for a fiber-based BB84 system \cite{boaron_secure_2018} and $1002~\si{\kilo\meter}$ using the twin-field approach \cite{liu_experimental_2023}.
Conversely, CV-QKD offers higher SKR for short lengths \cite{asif_seamless_2017, Wang:2024}.
Yet, due to loss sensitivity, the current record distance for fiber systems is limited to $203~\si{\kilo\meter}$ \cite{zhang_long-distance_2020}.
Nevertheless, compared with DV-QKD, CV systems can benefit from coherent detectors and widely accessible commercial integrated telecom components.
The former can work at room temperature and do not suffer from the typical Single Photon Detectors (SPDs) limitations, such as dead-time, commonly used in DV-QKD \cite{huang_dependency_2022}; a characteristic that positively contributes to the achievable SKR.
The second allows the exploitation of a widely consolidated industry, potentially enhancing the accessibility and economic viability of CV-QKD for broad adoption.

QKD initially emerged as a technology facilitating safe point-to-point communications.
However, progress in the field has now reached the stage where we are beginning to witness efforts to establish QKD networks involving multiple entities \cite{peev_secoqc_2009, sasaki_field_2011, aguado_engineering_2019}.
Such a quantum network implies the presence of multiple paths, including intermediary relay nodes, linking one end user to another.
Consequently, depending on the available routes and the channel's parameter connecting each node, hops may better use DV or CV protocols to maximize the overall SKR.
Rather than choosing between two dedicated transmitters, a more efficient strategy is to rely on a hybrid encoder design capable of switching between DV and CV \cite{grande_adaptable_2021}.
This approach offers the possibility of actively reconfiguring the network characteristics in a software-defined manner, thus increasing flexibility and improving the management and integration of QKD into the existing telecom infrastructure.
Although dynamically re-configurable and SDN-controlled quantum networks are a relatively recent development, their adoption has accelerated significantly in recent years {\cite{Chen2021, alia_dynamic_2022, Martin2024}} due to initiatives like the EuroQCI program.
A prime example of this trend is the MadQCI (Madrid EuroQCI network) \cite{Martin2024}, which successfully demonstrated dynamic {DV/CV reconfigurability} and SDN orchestration in a large-scale, deployed metropolitan network infrastructure.
Thus, we believe that the aforementioned factors represent key advantages in designing and implementing scalable QKD networks, particularly in dynamic environments.

In this work, we present the first hybrid encoder designed to be compatible with {well established state-of-the-art polarization-based DV and CV QKD protocols by reconfiguring its operating mode}.
In fact, all the other schemes previously proposed are limited to phase-encoded and time-bin DV systems \cite{grande_adaptable_2021}.
The system is realized only with fiber Commercial-Off-The-Shelf (COTS) telecom components and tested using two different setups for DV and CV, each paired with the respective receiver.
The paper is structured as follows: Section \ref{sec:Hybrid-encoder-principle} briefly outlines the working principles of the proposed hybrid encoder, Section \ref{sec:experimental-setup} describe the experimental setup adopted for CV and DV operations, while in Section \ref{sec:results} the results obtained.


\section{Hybrid QKD}\label{sec:Hybrid-encoder-principle}

\subsection{Applications of hybrid QKD encoders in quantum networks }
{Today's telecommunication systems rely on cryptographic procedures vulnerable to adversaries holding quantum computing capabilities.
This spurred research and industry to develop countermeasures like QKD, a solution devised to ensure Information Theoretic Security (ITS).}
However, the latter was mostly developed as a point-to-point connection requiring additional specialized infrastructure, which presents some challenges when integrating it into the current telecom grid.
{Moreover, such an approach encounters scaling issues for deploying QKD networks, as it is costly and increases the complexity of its management.}
Mitigating such issues is possible by relying on the Software Defined Network (SDN) model that is progressively embraced by telecom providers, given its ability to facilitate the integration of new services within a network infrastructure.
Since the success of QKD technology, either CV or DV, will depend on the ability to integrate it within the existing infrastructures, adopting the SDN method becomes appealing due to the aforementioned benefits.
Indeed, it should not impose the modification of each network device to establish a quantum link.
{Additionally, being SDNs centrally supervised, the routes connecting two endpoints can be reconfigured, allowing for the selection of the more appropriate operating mode, discrete or continuous, before starting the secret key exchange.
The latter is then maintained until it either completes successfully or aborts.
Only after the SDN can reconfigure the system, eventually to DV or CV.}
When paired with SDNs' inherited ability for real-time network monitoring, which includes quantum metrics for the case, it enables the optimization of the SKR between parties.
For instance, Hugues-Salas et al. in \cite{hugues-salas_monitoring_2019} demonstrated the ability to switch channels amid an attack, while Alia et al. in \cite{alia_dynamic_2022} showed link reorganization if the QBER exceeds a certain threshold.
The aforementioned flexibility and performance can improve by letting the controller adopt DV or CV QKD protocols for a given intermediate link based on the values of the observed quantum parameters.
For example, the controller can utilize CV-QKD for shorter connections as it typically offers higher transmission rates compared to DV-QKD, while preferring the latter for longer distances due to its ability to extend much further with current technology.
The method can also better adapt to the already deployed telecommunication infrastructure, which commonly employs multiple wavelengths for communication.
Indeed, in such an integrated scenario, quantum and classical channels will likely share the same propagating media.
{An heterogeneous network, capable of handling DV and CV links,} can provide the SDN controller with a broader range of options to enhance node-to-node communication and organize the communication considering the particular status of the link.

Within this context, a transmitter equipped with an encoder that can generate  quantum states {for both CV and DV QKD protocols} and easily switch between {them} well suits the task.
Moreover, such hybrid encoders may be a significantly cheaper alternative to two dedicated units while having a less complex yet compact and more adaptable design that can contribute to better system integration without compromising functionality.

\subsection{Hybrid CV-DV encoder}
\begin{figure*}
    \centering
    \includegraphics[width=0.86\linewidth]{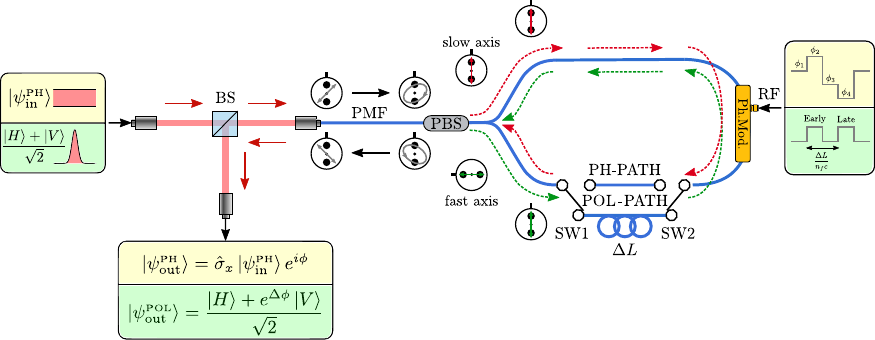}
    \caption{ 
        A schematic representation of our proposed hybrid encoder design.
        The device is set in DV or CV mode according to the optical path selected: POL-PATH for the former and PH-PATH for the second.
        In the DV configuration, the encoder is the standard iPOGNAC in which the source is a pulsed laser aligned in such a way that the light enters the system diagonally polarized.
        The latter vertical and horizontal components are spatially and temporally separated by a PBS and an optical delay line $\Delta L$ and undergo an early and late phase shift through the phase modulator.
        The electrical signals required to obtain such phase shifts are square pulses separated by $n_f \Delta L/c$, where $n_f$ is the PMF fiber slow-axis refractive index and $c$ is the speed of light in vacuum.
        In CV mode, instead, the laser source operates in CW (Continuous Wave), the $\Delta L$ is bypassed, and by properly setting the amplitude of the electrical pulses, M-PSK modulation constellations can be obtained.
    }
    \label{fig:ipognac_cv-dv}
\end{figure*}

The hybrid encoder design we introduce relies on the iPOGNAC \cite{avesani_stable_2020}, a polarization modulator that showed remarkable temporal stability due to a self-compensating Sagnac loop.
Such a technique gained attention in the QKD field \cite{qi_quantum_2006, roberts_patterning-effect-free_2018, Zhao:22, mandil_long-fiber_2024} due to its ability to counteract any drifts leading to enhanced stability and simplicity.
The iPOGNAC's operational principle can be understood from Figure (\ref{fig:ipognac_cv-dv}).
As can be seen, diagonally polarized light pulses are injected into the modulator.
The input Beam Splitter (BS) acts like a circulator guiding the transmitted photons to a Polarization Beam Splitter (PBS) through a Polarization Maintaining Fiber (PMF).
Then, the PBS outputs are connected to form an asymmetric Sagnac loop comprising a delay line and a phase modulator.
The former enables the phase modulator to apply a different phase shift to the horizontal and vertical polarization propagating through the Sagnac loop.
Indeed, the light entering the latter from the PBS vertical output travels clockwise. Upon exiting the loop, it emerges horizontally polarized with an additional \virgolette{early} phase shift $\phi_e$  induced by the phase modulator.
On the other hand, the light entering through the PBS horizontal output moves counterclockwise, gains a \virgolette{late} phase shift $\phi_l$, and exits vertically polarized.
The \virgolette{early} and \virgolette{late} light pulses are recombined at the PBS and travel backward toward the BS.
The transmitted component exits the iPOGNAC, up to a global phase, with a polarization state:
\begin{equation}\label{eq:dv_ipognac}
    \ket{\psi_{\text{out}}^\text{\tiny POL}} =  \dfrac{\ket{H} + e^{i  \left( \phi_l - \phi_e \right)} \ket{V}}{\sqrt{2}}\, .
\end{equation}
Notice that by adjusting the phases $\phi_e$ and $\phi_l$, it is possible to generate any balanced superposition of horizontal and vertical polarization states.

By removing the iPOGNAC's delay-line $\Delta L$, the Sagnac loop is now symmetric.
In this configuration, the same phase shift is applied simultaneously to both polarizations propagating through such a loop.
In a left-handed reference frame whose z-axis is directed toward the photon propagation direction, the PBS and Sagnac loop map $\ket{H}\rightarrow-\ket{V}$ and $\ket{V}\rightarrow \ket{H}$, which is described by the $i \hat{\sigma}_y$ operator.
If also the action of the BS is considered, then the device implements a $\hat{\sigma}_x$ operation up to a deterministic global phase.
As a result, regardless of the injected polarization state, the output is a fixed transformation of the input with an additional phase shift $\phi$ set by the phase modulator:
\begin{equation}
    \ket{\psi_{\text{out}}^\text{\tiny PH}} = \hat{\sigma}_x \ket{\psi_{\text{in}}} \, e^{i \phi}\, ,
\end{equation}
allowing thus the implementation of M-PSK (Phase Shift Keying) modulation schemes, where M is the discrete set of $\phi$ values used.
The geometry of the Sagnac loop ensures that, from the phase modulator perspective, both clockwise and counterclockwise components are always aligned with only one of its axis.
This prevents the phase of the input's $\ket{H}$ and $\ket{V}$ components from being unevenly modulated due to the modulator's polarization-dependent characteristics.
The aforementioned property is also a welcome feature since CV-QKD  systems usually employ phase modulators that exhibit unwanted polarization dependency that must be controlled and stabilized with additional equipment.

Since the iPOGNAC and the CV phase encoder differ only by the optical delay line from a hardware perspective, the two can be integrated into a single system.
The operational mode of this unified device can be set through two optical switches, SW1 and SW2, as illustrated in Figure (\ref{fig:ipognac_cv-dv}).
Indeed, the latter enables $\Delta L$ to be bypassed.
Switches SW1 and SW2 can be implemented in different ways.
However, since the operating mode is established before starting the QKD protocol and maintained until it is completed successfully or aborted, we do not foresee the need for high performance in terms of speed and latency.
Hence, we believe MEMS optical switches will suffice for this particular case.
They are straightforward to use, relatively inexpensive, reliable, and have low insertion loss.
Furthermore, the latter can also be electrically controllable, which implies that transitioning from DV to CV, or the reverse, can be achieved remotely with ease, in real-time, and without significant effort.


\section{Experimental setup}\label{sec:experimental-setup}
The capabilities of our proposed hybrid scheme were demonstrated using a DV and CV setup, each of which had its corresponding receiver.
Given the nature of these experiments and the fact that detectors capable of working for both DV and CV protocols are missing,
besides early proposals that may hint to such a direction \cite{Qi:2021,sidhu:2024}, we avoided testing the switching mechanism.
However, future work will focus on implementing such a mechanism with MEMS technology and assessing it with appropriate experiments.

\subsection{CV setup}
\begin{figure*}
    \centering
    \includegraphics[scale=0.88]{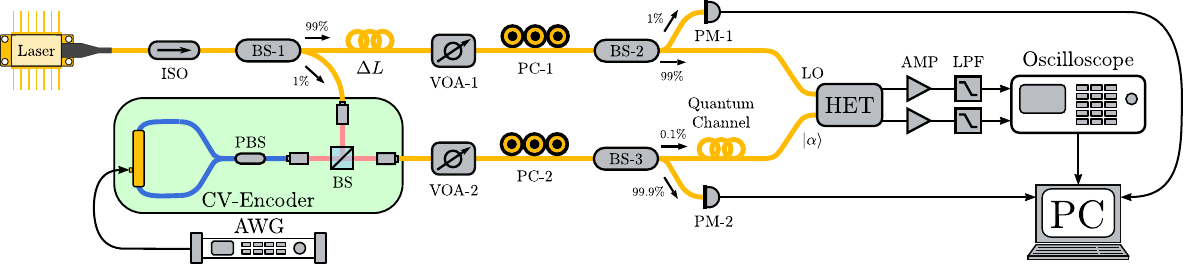}
    \caption{Schematic representation of the CV experimental setup used.
    The latter consists of a $1550~\si{\nano\meter}$ Continuous Wave (CW) laser, the output of which is divided by a 99:1 beam-splitter (BS-1).
    The majority of such (99\%) is directed towards a heterodyne receiver (HET) acting, thus, as a Local Oscillator (LO).
    The rest (1\%) is sent to the CV encoder for CV quantum state phase modulation and then routed, through a fiber quantum channel, to the heterodyne's signal input port.
    Subsequently, the heterodyne measurements are digitized by an oscilloscope and delivered to a PC for data processing.
    Notice that additional optical components are present between the 99\% BS-1 arm and the HET, as well as between the CV encoder and the HET.
    The former are required to characterize the heterodyne receiver (such as photodiode linearity and clearance), while the latter attenuate the modulated coherent state generated by the encoder to the quantum level.
    Finally, the image provides a color-coded representation of the fibers used: blue for single-mode polarization-maintaining and yellow for single-mode non-polarization-maintaining.}
    \label{fig:cv-setup}
\end{figure*}
A schematic representation of the experimental setup used to test our hybrid encoder in CV mode is represented in Figure (\ref{fig:cv-setup}).
As can be seen, a 99:1 Beam Splitter (BS-1) divides the light of a single-mode continuous-wave $1550~\si{\nano\meter}$ laser with a line-width of $\sim 100~\si{\kilo\hertz}$.
An optical isolator (ISO) is positioned before BS-1 to prevent back-reflected light from causing laser instabilities.
Then, $99\%$ of the light from BS-1 is directed to an integrated COTS heterodyne receiver with approximately $13~\si{dB}$ of clearance, serving as a Local Oscillator (LO).
Before entering the receiver, the light passes through a Variable Optical Attenuator (VOA-1), a Polarization Controller (PC-1), and a 99:1 Beam-Splitter (BS-2).
This splitter sends $1\%$ of the light to a monitoring Power Meter (PM-1) while the remaining $99\%$ to the photonic chip.
The power meter PM-1 and the VOA-1 are required only for calibrating the heterodyne receiver, a procedure in which we test the linearity of its photodiodes and estimate the parameters to convert the measured heterodyne signals from volt units into shot-noise units.
The polarization controller, instead, is necessary to maximize the amount of optical power coupled into the chip due to the LO inlet being polarization-sensitive.
Conversely, the BS-1 $1\%$ branch sends the light to the hybrid encoder configured to work in CV.
The light from this encoder then travels through VOA-2, the polarization controller PC-2, and the 99.9:0.1 BS-3.
Given the dual-polarization feature of the heterodyne receiver, PC-2 aligns the light polarization state to ensure that only one of the two polarization-dependent heterodyne is used.
Moreover, with this cascade we can attenuate the incoming light to the required quantum level and monitor such with the PM-2 power meter at the $99.9\%$ branch of BS-3.
The BS-3 $0.1\%$ branch is subsequently connected to the heterodyne's signal input port through a quantum channel.
Notice that the signal and LO paths share the same $1550~\si{\nano\meter}$ laser and maintain identical lengths, with additional $\Delta L$ fiber at the LO branch, to minimize phase noise and instabilities at the receiver.
Since our objective is to evaluate our CV encoding scheme, this approach eases the experimental implementation compared to the more secure Local-Local oscillator technique \cite{soh_self-referenced_2015, qi_generating_2015}.
After detection, the electrical signals from the integrated chip are amplified using two RF amplifiers and subsequently filtered by two RF low-pass filters.
Each RF component has a bandwidth of $500~\si{\mega\hertz}$.
Then, these signals are digitized by a $4~\si{\giga\hertz}$ oscilloscope with a resolution of 8-bit and a sampling rate of $25~\si{GS/s}$.
Finally, the digitized data is streamed to a computer for further offline analysis.
The latter procedure accounts for constellation reconstruction and estimation of the main encoder parameters for CV-QKD, i.e. electronic noise, excess noise, and transmittance.
To test the performance of our CV encoder, we used the $\phi$-phase shifts $\pi/2$, $3\pi/2$, $-3\pi/2$ and $-\pi/2$.
The electrical control signals were generated by a 16-bit Arbitrary Waveform Generator (AWG) operating at a symbol rate of $50~\si{MBaud}$.
To simplify post-processing, we synced the oscilloscope's data acquisition with the AWG via a reference clock.

\subsection{DV setup}
\begin{figure*}
    \centering
    \includegraphics[width=0.98\linewidth]{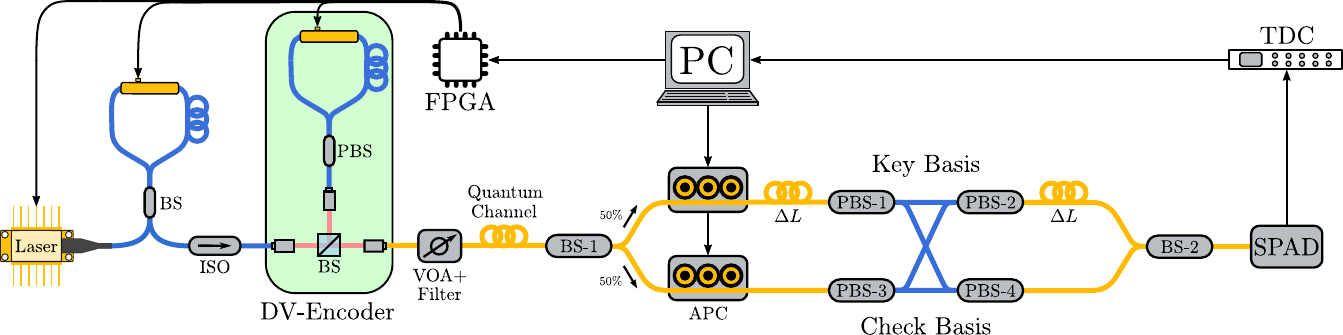}
    \caption{The system employs a $1550~\si{\nano\meter}$ pulsed laser, modulated by a Sagnac interferometer and iPOGNAC for decoy and quantum states $\ket{D}$, $\ket{R}$, $\ket{L}$ generation respectively.
    To guarantee that the iPOGNAC outputs $\ket{D}$, $\ket{R}$, $\ket{L}$, the input light must be $\ket{D}$, a condition achieved by carefully aligning such polarization modulator with the Sagnac, and the phase modulator must be operated with the appropriate electrical signals.
    The light is then attenuated to the single photon level by a Variable Optical Attenuator (VOA), filtered by a narrow-band optical filter centered at $1550.12~\si{\nano\meter}$ to reject noise, and directed, through a fiber quantum channel, to a time-multiplexed receiver with a Single Photon Avalanche Detector (SPAD) and two Automatic Polarization Controller (APDs) for QKD measurement.
    Finally, a time-tagger (TDC) records the SPAD signals for PC analysis.
    Notice that the signals controlling the Laser, Sagnac, and iPOGNAC are managed by an FPGA to ensure the correct timings.
    Moreover, the image provides a color-coded representation of the fibers used: blue for single-mode polarization-maintaining and yellow for single-mode non-polarization-maintaining fibers.}
    \label{fig:dv-setup}
\end{figure*}
The experimental DV setup used, as shown in Figure (\ref{fig:dv-setup}), is similar to the one described in \cite{avesani_resource-effective_2021} for the transmitter, and to the one detailed in \cite{avesani_deployment-ready_2022} for the receiver.
The source is a $1550~\si{\nano\meter}$ gain-switched laser emitting phase randomized pulses with a temporal width of $270~\si{\pico\second}$, measured at full-width-half-maximum, with a repetition rate of $50~\si{\mega\hertz}$.
The pulse amplitude is then modulated with a Sagnac interferometer \cite{roberts_patterning-effect-free_2018}, which includes a 70:30 BS, a phase modulator, and an optical delay line of one meter.
This Sagnac modulator is employed to generate the decoy states.
Before entering the iPOGNAC, the polarization of the light pulse is rotated to guarantee that it enters diagonally polarized.
We obtain such by rotating the iPOGNAC's input fiber collimator.
Subsequently, the pulse polarization is adjusted using an iPOGNAC to produce the state $\ket{D}$ (diagonal), $\ket{R}=\left( \ket{H}-i\ket{V} \right)/\sqrt{2}$ (circular right-handed), and $\ket{L}=\left( \ket{H}+i\ket{V} \right)/\sqrt{2}$ (circular left-handed) by setting $\phi_l-\phi_e$ respectively equal to $0$, $-\pi/2$, and $+\pi/2$.
Before being sent to the receiver through a fiber quantum channel with $\sim 6.6~\si{dB}$ of losses, these states are attenuated to the single-photon level by a calibrated VOA and filtered with a narrow-band optical filter centered at $1550.12~\si{\nano\meter}$ to reduce noise.
The receiver then detects the incoming states by randomly alternating between the basis sets $\{\ket{D},\ket{A}\}$ (check basis) and $\{\ket{R},\ket{L}\}$ (key basis), each with an equal probability of selection.
Such selection is passively realized using a 50:50 BS, and each projective measurement is implemented with an automatic polarization controller and a PBS.
Moreover, a time and polarization multiplexing scheme performs the measurements using only one InGaAs/InP Single-Photon Avalanche Detector (SPAD).
Finally, the photon's time-of-arrival is recorded by a Time-to-Digital-Converter (TDC), and the polarization and time reference frame synchronization is performed continuously and automatically adjusted by the Qubit4Sync algorithm by using a few qubits from the quantum communication \cite{calderaro_fast_2020}.


\section{Results}\label{sec:results}
\begin{figure}
	\centering
	\includegraphics[width=1\linewidth]{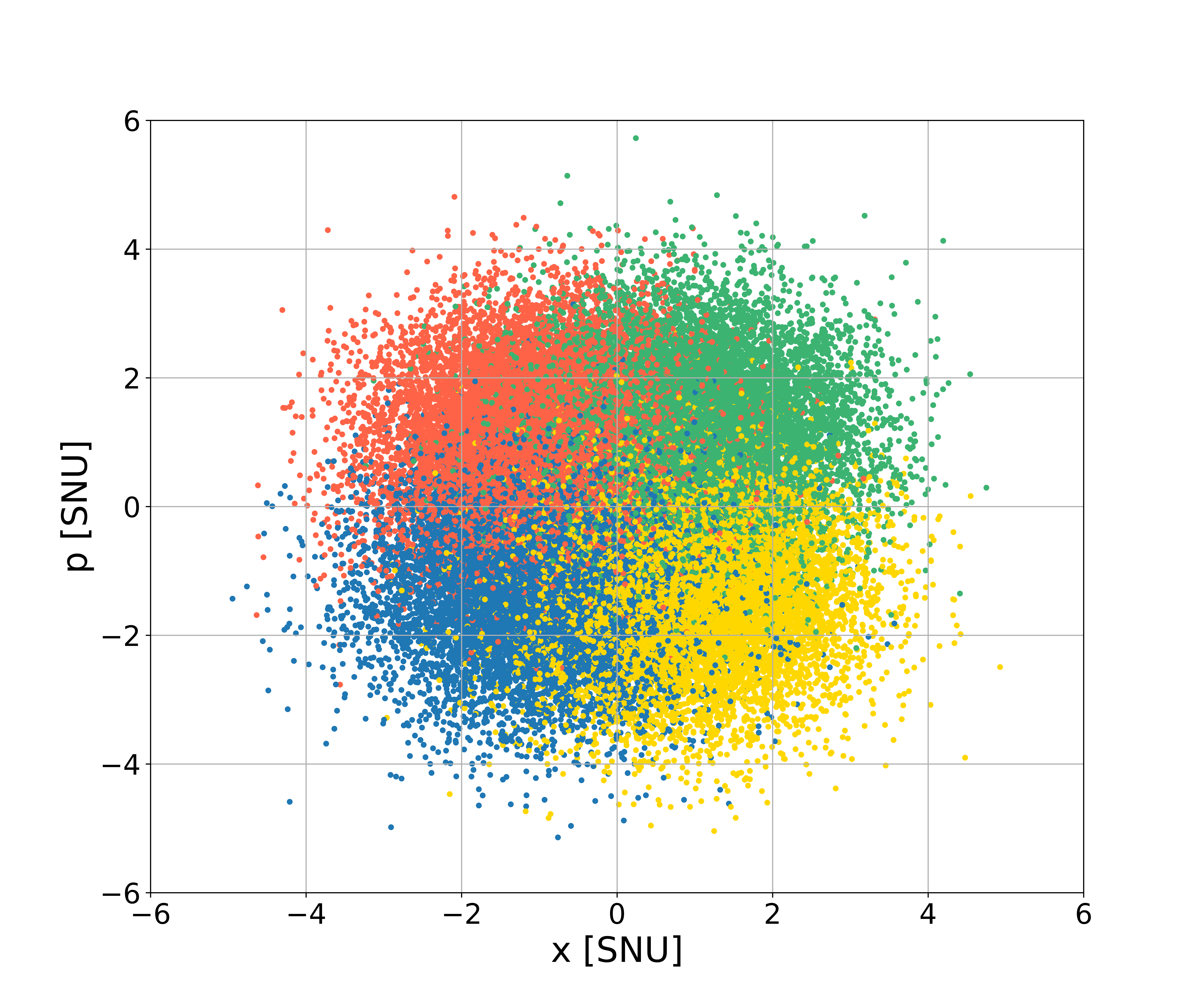}
	\caption{The picture showcases an example of QPSK constellation obtained with our CV-Encoder operating at a symbol rate of $50~\si{MBaud}$ and with Alice's variance $\text{V}_\text{A}$ set to $\sim 12.4~\si{SNU}$.}
	\label{fig:symbols-qpsk}
\end{figure}

In the experiment with the CV phase encoder, we repeatedly transmitted four coherent states that formed a QPSK constellation.
Figure (\ref{fig:symbols-qpsk}) depicts an example of the received constellation when Alice's modulation variance is approximately $12.4~\si{SNU}$.
Such a value is not suited for CV-QKD purposes.
However, it allows us to resolve the constellation enough to showcase the encoder's modulation capabilities.
For CV-QKD applications, the employed variance $\text {V}_\text{A}$ is detailed in Table (\ref{tab:results-parameter-estimation}).

Table (\ref{tab:results-parameter-estimation}) reports the parameters estimated for our phase encoder.
\begin{table}
    \centering
    \renewcommand{\arraystretch}{1.2} 
    \begin{tabular}{cccc}
        \hline \addlinespace
        \textbf{Parameter} & \textbf{Symbol} & \textbf{Value} & \textbf{Units} \\ \hline \addlinespace
        \makecell{Electronic noise \\ variance} & $\text{V}_{\text{el}}$ & $0.081$ & SNU \\ \hline
        Receiver losses & $\eta$ & $0.30$ & \\ \hline
        Alice's variance & $\text{V}_{\text{A}}$ & $0.45$ & SNU \\ \hline
        Channel transmittance & $\text{T}$  & $0.72$ & \\ \hline
        Excess noise & $\xi_{\text{A}}$  & $0.012$ & SNU\\ \hline
        SDP Asymptotic & & & bits/ \\
        Secret Key Rate & $\text{SKR}$ & $0.021$ & symbol \\ \hline
        LC Asymptotic & & & bits/ \\
        Secret Key Rate & $\text{SKR}$ & $0.026$ & symbol \\ \hline
    \end{tabular}
    \caption{Values of the parameter estimation using the CV-Encoder with $50~\si{MBaud}$ QPSK modulation and setting the information reconciliation efficiency to $95\%$. 
    The results are obtained by analyzing $30~\si{samples}$, each with $50\, 000~\si{symbols}$.}
    \label{tab:results-parameter-estimation}
\end{table}
Assuming a trusted detector scenario, the heterodyne receiver was calibrated estimating its electric noise $\text{V}_\text{el}$ and losses $\eta$ (the product of the efficiency of its photodetectors and transmittance) which resulted approximately $0.081~\si{SNU}$ and $0.27$ respectively.
The modulation variance $\text{V}_\text{A}$ of the symbols transmitted by Alice was measured by the power meter PM-2 giving the value $\sim0.45~\si{SNU}$ (we recall that such quantity is related to the mean photon number $\mu$ of the generated states as $\text{V}_\text{A}=2\mu$).
The excess noise parameter $\xi_\text{A}$ and channel transmittance $\text{T}$ were obtained by the following relations:
\begin{align}
    & \expval{\text{X}_\text{A} \text{X}_\text{B}} = \sqrt{\frac{\eta \text{T}}{2}} \text{V}_\text{A} \\
    &\text{V}_\text{B} = 1 + \text{V}_{\text{el}} + \frac{\eta \text{T}}{2} \text{V}_\text{A} + \frac{\eta \text{T}}{2} \xi_\text{A}
\end{align}
with $\expval{\text{X}_\text{A} \text{X}_\text{B}}$ the covariance matrix elements and $\text{V}_\text{B}$ Bob's variance estimated by analyzing a dataset of $1.5\cdot 10^6$ symbols acquired with the oscilloscope.
The ideal asymptotic SKR obtained with our device is approximately $0.021~\si{bits/symbol}$, which leads, given a symbol-rate of $50~\si{MBaud}$, to a bit-rate of $\sim 1.1~\si{Mbps}$.
To compute the latter, we use the well-known Devetak-Winter bound \cite{devetak_distillation_2005} $\text{SKR}=\beta \, \text{I}_{\text{AB}}-\chi_{\text{E}}$, where $\beta$ set at a value of $95\%$ (typically employed in the literature \cite{denys_explicit_2021,wang_sub-gbps_2022}) is the error correction efficiency, $\text{I}_{\text{AB}}$ is the mutual information between Alice and Bob, and $\chi_\text{E}$ the Holevo bound on the information between Bob and Eve in the reverse reconciliation scenario.
To compute the Holevo bound $\chi_\text{E}$, we relied on a recent security proof based on a Semi-Definite Programming (SDP) method outlined in \cite{ghorai_asymptotic_2019,denys_explicit_2021} which directly considers discrete modulations schemes including QPSK.
However, the approach does not consider hardware non-idealities like the electronic noise $\text{V}_{\text{el}}$ and the detector's efficiency $\eta$.
For completeness, we also report the performance obtained with the Linear Channel (LC) model, almost $0.026~\si{bits/symbol}$ and $\sim 1.3~\si{Mbps}$ given the $50~\si{MBaud}$ rate, which takes into account the aforementioned non-idealities, although it restricts the possible attacks that the eavesdropper can perform.
{In both cases, the tolerable excess noise for QPSK is very low, limiting the SKR and reachable distances. Improving the latter requires larger constellation schemes \cite{zhao_unidimensional_2020, denys_explicit_2021}.}

\begin{figure}
    \centering
    \includegraphics[width=1\linewidth]{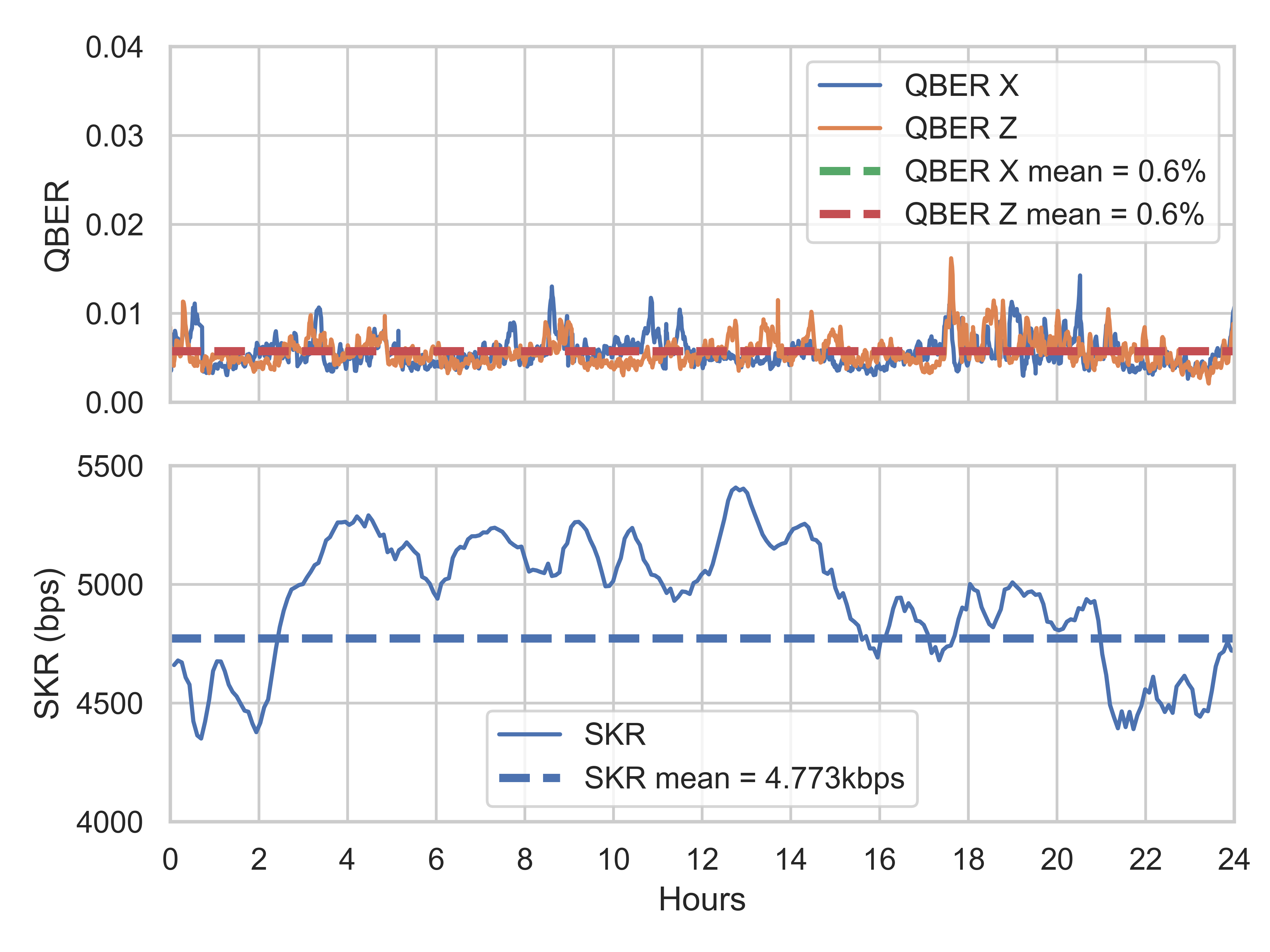}
    \caption{The picture illustrates the performance achieved during $24~\si{hours}$ of continuous operation with our hybrid encoder in DV mode.
    The upper panel displays the QBER for both the key (Z) and check (X) basis, averaging $0.6\%$ in each case.
    The lower panel shows the associated finite-size SKR, which averages $4.7~\si{kbps}$.}
    \label{fig:dv-result}
\end{figure}

For the DV encoder, we estimated the QBER and SKR when implementing the efficient three-states one-decoy protocol.
The results are shown in Figure (\ref{fig:dv-result}).
As can be seen, the finite-size SKR shows very high stability during $24~\si{hours}$ of data acquisition, with an average of $4.7~\si{kbps}$.
This is further demonstrated by the intrinsic QBER plot, which remains very low with an average of $0.6\%$ for both bases.
The results further confirm the excellent performance of the iPOGNC as a polarization encoder in terms of QBER and long term stability.


\section{Conclusions}
Hybrid QKD devices are powerful tools for managing a complex quantum network.
When paired with the functionalities provided by an SDN, these systems can enable a more flexible and optimized allocation of its resources.
Such versatility is largely due to their ability to toggle between DV and CV protocols, which is fundamental for implementing these networks.
{In this work, we introduce a hybrid QKD encoder designed to support well-established state-of-the-art CV and polarization-based DV protocols by switching operating modes upon an external control signal.
To the best of our knowledge, this device is the first of its kind.}
Our approach leverages an iPOGNAC for DV, where the polarization encoder's Sagnac loop is asymmetric due to an optical delay line, and for CV applications, in which the loop is symmetric as the previously mentioned delay is bypassed.
We remark that since we are exploiting a Sagnac loop, all perturbations are canceled.
For DV-QKD, this improves the system's QBER.
Indeed, we obtained a low value of $\sim 0.6\%$ for the latter over $24~\si{hours}$, which led to a finite size SKR of $\sim 4.7~\si{kbps}$.
For the CV scenario, alongside improved stability, the loop's structure also makes the encoder polarization-insensitive, and we were able to achieve an asymptotic SKR of almost $0.021~\si{bits/symbol}$, which translates to a bit-rate of $\sim 1.1~\si{Mbps}$ with the symbol-rate of $50~\si{MBaud}$ used.
Concerning the latter insensitivity, due to the light horizontal and vertical components always being aligned with only one of the phase modulator's principal axes, we would like to highlight that this is a welcome feature for CV-QKD.
In fact, such systems typically adopt modulators that exhibit unwanted polarization dependency.
To ensure proper functioning, they must be controlled and stabilized with extra equipment that adds system complexity.
In light of such, our scheme also simplifies the realization of a CV state encoder by avoiding the need for the aforementioned additional equipment.

Furthermore, we believe that the aforementioned compatibility of our device with polarization-based DV-QKD makes our scheme an attractive solution for free-space links for which polarization is usually employed, given its enhanced robustness considering the properties of the free-space channel itself.
It is worth noting that free-space channels include satellite-to-satellite and satellite-to-ground connections, which are of great interest both for research and industry.
On top of this, there is a growing interest in developing CV-QKD systems for satellite applications.
Within such a context, our scheme can provide additional benefits other than integrating a DV and CV state encoder due to its lower complexity compared to two dedicated modules, such as a reduction in size, power consumption, and cost over two dedicated encoder modules due to its lower complexity.

In conclusion, this work introduces a hybrid {quantum state} encoder compatible with CV and polarization-based DV QKD protocols built entirely from COTS components.
Furthermore, it features high stability alongside polarization insensitivity, and by integrating supplementary external components, it can also support additional phase-based DV protocols.
We believe that our design represents an advancement in the realization of flexible and re-configurable nodes for the quantum networks of the future in which simplicity and compactness also matter.

\section*{Funding}
This work was supported by European Union’s Horizon Europe research and innovation program under the project Quantum Secure Networks Partnership (QSNP), grant agreement No 101114043.

\begin{acknowledgments}
The authors would like to thank Dr. Matteo Schiavon and Dr. Yoann Piétri for the useful discussions.
The authors also thank the CloudVeneto facility for computational resources.
The authors would like to thank also the European Union’s Horizon Europe research and innovation programme under the project QUANGO (grant agreement No 101004341) and the project “Quantum Secure Networks Partnership” (QSNP, grant agreement No 101114043) for providing preliminary research basis for this work. Views and opinions expressed are however those of the author(s) only and do not necessarily reflect those of the European Union or European Commission-EU. Neither the European Union nor the granting authority can be held responsible for them.
\end{acknowledgments}
\section*{Disclosure}
The authors declare no conflicts of interest.

\section*{Data availability}
Data supporting the results presented in this paper are
available from the authors upon reasonable request.

\end{document}